\documentstyle[prl,aps,psfig]{revtex}
\begin{document}
\draft
%
\preprint{\footnotesize Revised \today}
\twocolumn[
\title{
Field-Induced Staggered Magnetic Order in La$_2$NiO$_{4.133}$}
\author{J. M. Tranquada,$^1$ P. Wochner,$^1$ A. R. Moodenbaugh,$^1$ and D. J.
Buttrey$^2$}
\address{$^1$Physics Department, Brookhaven National Laboratory, Upton, 
New York 11973}
\address{$^2$Department of Chemical Engineering, University of Delaware,
Newark, Delaware 19716}
\date{November 8, 1996}
\maketitle
\widetext
\advance\leftskip by 57pt
\advance\rightskip by 57pt
\begin{abstract}
At low temperature the holes doped into the NiO$_2$ planes of
La$_2$NiO$_{4.133}$ by the excess oxygen collect in diagonal stripes
that separate narrow antiferromagnetic domains.  The
magnetic order drops abruptly to zero at $T_m=110.5$~K, but charge order remains
with a period of $\frac32a$. We show that application of a magnetic field
in the regime $T>T_m$ induces staggered magnetic order of period
$3a$ due to the net magnetic moment of the high-temperature bond-centered
stripes, together with the odd number of Ni spins across an antiferromagnetic
domain.
\end{abstract}
]
\narrowtext

It is now experimentally established that holes doped into the NiO$_2$ planes of
La$_2$NiO$_4$ tend to order in a periodic structure consisting of parallel
charge stripes \cite{chen93,tran94a,sach95}.  The
segregation of the holes into charge stripes leaves intervening regions that are
essentially undoped.  The magnetic moments of the Ni ions in these regions are
correlated antiferromagnetically \cite{hayd92,yama94}.  Neighboring
antiferromagnetic domains, separated by a charge stripe, have an antiphase
relationship; that is, the phase of the magnetic order shifts by $\pi$ on
crossing a charged domain wall \cite{tran94a,sach95}.  Evidence
for related stripe correlations has been found in hole-doped La$_2$CuO$_4$
\cite{cheo91,thur92,tran95a}, and there are indications that
dynamical stripe correlations have a connection with the superconductivity found
in the layered cuprates \cite{tran96c}.

The problem of stripe order in a doped two-dimensional antiferromagnet has
received considerable attention from theorists
\cite{zaan89,zaan94,low94,vier94,haas95,tsun95,seib96,sale96,naya96,whit96}. 
One feature of theoretical interest concerns the alignment of the charge stripes
with the lattice.  In particular, one would like to know whether the domain
walls are centered on rows of metal atoms (site-centered stripes) or on rows of
oxygens (bond-centered stripes).  This alignment is difficult to determine in a
standard scattering experiment because of the loss of the phase information
carried by the scattered beam.  In the present paper, we determine the stripe
alignment in a crystal of La$_2$NiO$_{4+\delta}$ through an unusual effect, in
which a staggered magnetization is induced in a paramagnetic phase by the
application of a uniform magnetic field.  This effect can be understood as a
ferrimagnetic response associated with the ferromagnetic nature of bond-centered
stripes in the high-temperature charge-ordered phase.

The particular crystal of La$_2$NiO$_{4+\delta}$ studied has an oxygen excess
of $\delta=\frac2{15}=0.133$, and a detailed characterization of the charge and
spin order will be presented elsewhere \cite{woch96}.  The oxygen interstitial
order is the same as that in samples with a nominal $\delta=0.125$ studied
previously \cite{tran94a}, but we believe that $\delta=\frac2{15}$
corresponds to the optimal interstitial concentration for this particular
phase.  The magnetic Bragg peaks measured on the present crystal are sharper
than those observed in our own previous work, allowing better sensitivity to
intrinsic properties.

To provide a context for understanding our new results, it is first necessary to
review some details concerning the
previously determined order \cite{tran94a,woch96}.  We consider a unit
cell of size $\sqrt{2}a_t\times\sqrt{2}a_t\times c$ relative to the tetragonal
unit cell of the K$_2$NiF$_4$ structure.  Antiferromagnetic order within an
NiO$_2$ plane would then be characterized by the modulation wave vector 
${\bf Q}_{\rm AF}=(1,0,0)$, where the components are in units of 
$({2\pi\over a},{2\pi\over a},{2\pi\over c})$.  The charge order is
characterized by the wave vector $(2\epsilon,0,0)$ \cite{note2}, with the
average distance between domain walls in real space equal to $a/2\epsilon$; the
magnetic modulation is $(\epsilon,0,0)$ with respect to ${\bf Q}_{\rm AF}$. 
(Note that, relative to the simple square lattice of an NiO$_2$ plane, the
stripes run diagonally.)  For $\delta=\frac2{15}$, charge order occurs at
$T\lesssim220$~K with $\epsilon=\frac13$, while magnetic order appears abruptly
at $T_m=110.5$~K with a concomitant jump in $\epsilon$ to 0.295.  The modulation
parameter $\epsilon$ continues to decrease as the temperature is lowered below
$T_m$.

At 10~K we find $\epsilon=\frac5{18}=0.278$ \cite{woch96}.  This is close to the
value 0.266 ($=2\delta$) that one would expect to find if there were exactly one
hole per site along a charge stripe, as suggested by the calculations of Zaanen
and Littlewood \cite{zaan94}.  The fact that $\epsilon$ increases with
temperature indicates that the density of stripes becomes greater as the
magnetic order parameter is reduced.  [Note that, with a fixed hole
concentration, the hole density (per Ni site) within a stripe, given by
$2\delta/\epsilon$, must correspondingly decrease.]  In principle, the change in
density could be accommodated with a single type of stripe (either site- or
bond-centered).  For example, near 100~K where
$\epsilon$ locks into a value of $\frac27$, the observed wave vector could be
explained by alternating stripe spacings of $\frac32a$ and $2a$.  In terms of
the magnetic order, the positions of the domain walls would be far from the
nodes of a sinusoidal modulation with the same wave vector, and hence one would
expect significant magnetic harmonics at $3\epsilon$ and $5\epsilon$.  On the
other hand, if the stripes alternate between site- and bond-centered positions,
then they have a uniform spacing of $\frac74a$, resulting in much weaker
harmonics.  Quantitative analysis of experimentally observed harmonic
intensities supports the latter model.  Hence, it appears that there is a mixture
of site- and bond-centered stripes whose respective densities change with
temperature; however, we do not know experimentally which type dominates at low
temperature. 

Now let us consider the situation when $\epsilon=\frac13$.  In this state the
charge is still ordered, but the Ni spins are only dynamically correlated.  The
wave vector is such that the stripes must all be either site-centered or
bond-centered; these two possibilities are illustrated in Fig.~1.  In the
figure we have also indicated the spin arrangements one might find if a
``snapshot'' were taken.  For case (a), the antiferromagnetic domains are just
2 spins wide, and each spin has an antiparallel partner.  In case (b), the
domains are 3 spins wide, and an uncompensated moment may appear.  The magnetic
moments on sites adjacent to a charge stripe are likely to be reduced in
magnitude compared to those in the middle of a domain, but perfect compensation
would be a suprising coincidence.  Furthermore, every domain contains two up
spins for each down spin, so that this spin configuration should exhibit a
ferrimagnetic response.  Note that in both cases (a) and (b) the phase of the
antiferromagnetic order shifts by $\pi$ on crossing a domain wall; however, for
a bond-centered stripe the adjacent spins are ferromagnetically aligned,
whereas for a site-centered stripe adjacent spins are antiparallel.

Magnetization measurements by Yamada {\it et al.} \cite{yama94} provided the
first indication of a ferrimagnetic response in the paramagnetic phase.  They 
found a sharp peak in the
magnetization at the magnetic ordering temperature when the
magnetic field is applied parallel to the NiO$_2$ planes.  (An example of such a
measurement on a piece of our crystal is shown in Fig.~2.)  They also showed that
the peak disappears when the field is applied along the $c$ axis, perpendicular
to the planes.  (We have checked that there is no significant dependence of the
magnetization on the direction of the field within the plane, so the spins are
apparently XY-like.) The peak was attributed to the response of Ni spins with
spiral correlations within the planes, by analogy with the response associated
with out-of-plane spin canting that occurs in La$_2$CuO$_4$
\cite{thio88} and La$_2$NiO$_4$ \cite{yama92}; however, we have shown
elsewhere \cite{tran94a} that, in the magnetically ordered phase, the spins are
essentially collinear and oriented parallel to the stripes, which is
inconsistent with spiral order.  

In terms of the stripe model, the peak in the magnetization for $T=T_m^+$ appears
to be evidence for bond-centered ferromagnetic domain walls.  If this picture is
correct, then it should be possible to induce a staggered magnetization by
applying a uniform magnetic field.  To test this model, we performed a neutron
diffraction experiment on a piece of the same crystal of La$_2$NiO$_{4.133}$
that we have characterized in detail elsewhere \cite{woch96}.  The crystal was
mounted in a flow cryostat that was placed in a vertical-field
superconducting magnet.  The [010] axis of the crystal was aligned parallel
($<1^\circ$ error) to the magnetic field, both of which were perpendicular to
the scattering plane.  Elastic scattering measurements were performed using
5-meV neutrons at the H9A triple-axis spectrometer, located at the High Flux
Beam Reactor, Brookhaven National Laboratory.  Most of the scans were performed
along ${\bf Q}=(h,0,1)$, through magnetic peaks at $h=1-\epsilon$.

Representative scans are shown in Fig.~3.  In the scans of Fig.~3(a), measured
at $T=111\mbox{~\rm K}$ ($>T_m$), there is no magnetic peak in zero field
\cite{note3}, but a clear peak at $h=\frac23$ ($\epsilon=\frac13$) appears in a
field of 6~T.  The staggered magnetization (proportional to the square root of
the intensity) varies linearly with the applied field.  On cooling just below
$T_m$ [Fig.~3(b)], a zero-field peak appears at $h=0.705$ ($\epsilon=0.295$).  
As the field is raised from zero, the $h=0.705$ peak decreases in
intensity while the
$h=\frac23$ peak grows.  Rather than a smooth shift of the peak, there is a
coexistence of the two wave vectors.  When the applied field reaches 6~T, all
of the intensity is in the $h=\frac23$ peak.  By the time that the sample has
been cooled to 108~K, the 6-T field has essentially no effect on the magnetic
order [Fig.~3(c)]. Note that the width of the
$h=\frac23$ peak is resolution limited, both along
$h$ and in the transverse direction, along $l$ (not shown).  The $h$-widths of
the incommensurate peaks are slightly broader, indicating a small amount of
disorder in the stripe spacing.

Figure~4 shows the temperature dependence of the magnetic peak
intensity.  In zero field the magnetic peak intensity grows rapidly below the
first-order transition at 110.5~K.   In a 6-T field the $\epsilon=\frac13$
peak is observable over a wide temperature range $\gtrsim110$~K.  The inset
shows the logarithm of the normalized peak intensity versus temperature at 6~T. 
Since the intensity is proportional to the square of the staggered magnetization,
$M$, the linear variation (denoted by the fitted line) indicates that
\begin{equation}
  M^2 \sim e^{-2T/T_0},  
\end{equation}
with $T_0=20\pm5$~K.  The exponential dependence on temperature is similar to
what one would get from a Debye-Waller factor, suggesting that the
magnetization is limited by fluctuations of the correlated spins about the
stripe-ordered state.  (We have not tested for the $Q$ dependence that one
would expect for a Debye-Waller factor.)  Also shown in the inset is the
temperature dependence of a charge-order peak intensity, obtained in a separate
measurement in zero field
\cite{woch96}.  In the temperature range where the field-induced magnetic peak
is observed, the charge-order intensity is fairly constant; however, it is
somewhat surprising that, at higher temperatures, the charge-order
intensity also decreases exponentially with temperature.  The fit shown in the
inset of Fig.~4 corresponds to $T_0=67\pm5$~K.  This unusual behavior might
indicate that the charge correlations are fluctuating about the commensurate
lattice potential caused by the ordered interstitial oxygens.  For
$\epsilon=\frac13$, the charge-order wave vector is equal to the second
harmonic of one of the two interstitial-order wave vectors \cite{tran94a}.

The observation of field-induced magnetic scattering corresponding to
$\epsilon=\frac13$ is direct evidence that the high-temperature domain walls
are bond-centered.  For $T<T_m$ the density of stripes decreases, and the
stripes become increasingly site-centered.  The density of bond-centered
stripes (per diagonal row of Ni sites) is equal to $4\epsilon-1$, and can be
calculated using the results for $\epsilon$ from
Ref.~\cite{woch96}.  The bond-centered stripe density is shown by the squares in
Fig.~2.  A comparison of this density with the
magnetization shows that corresponding structures are found in the temperature
range 90--110~K.  This indicates that there is a bulk response from the
ferromagnetic domain walls even in the ordered state.  Note that there is no net
macroscopic ferrimagnetism in the ordered state because, with
$\epsilon<\frac13$, the ferromagnetic domain walls are no longer all in phase
with one another.  The drop in the magnetization at $T<10$~K suggests another 
shift in $\epsilon$ below the minimum temperature studied by neutron diffraction
\cite{woch96}.

What are the implications of these results for stripe correlations in the
cuprates?  
One significant difference in the cuprates is that the domain walls are
``vertical'' (or ``horizontal'') instead of diagonal within a square
lattice \cite{tran95a}.  As a result, the Cu spins next to a domain wall
alternate in direction as one moves along the wall.  With one hole per site
along a wall, all spin correlations would remain antiferromagnetic; however, for
doped La$_2$CuO$_4$ the hole concentration is $\approx\frac12$ hole per site
along a charge stripe.  For a bond-centered domain wall, it is possible to
imagine a configuration that yields a net magnetic moment.  It would be
interesting to see whether such a configuration can be detected in a real
compound.

We gratefully acknowledge helpful discussions with D. E. Cox, V. J. Emery, S.
A. Kivelson, and G. Shirane.  Work at Brookhaven was carried out under Contract
No.\ DE-AC02-76CH00016, Division of Materials Sciences, U.S. Department of
Energy.


\begin{thebibliography}{10}
\bibitem{note2}
We have previously reported that the wave vector for the charge modulation is
$(2\epsilon,0,1)$.  More recent results show that the charge modulations in the
two nearest-neighbor NiO$_2$ layers are not related by symmetry.  Instead, the
modulation phases for a pair of layers are related by the basis vectors
$(0,0,0)$ and $(a,0,\frac{c}2)$.
\bibitem{note3}
Charge-order scattering is also allowed at $(\frac23,0,1)$. Neutrons detect the
presence of charge order via the induced atomic displacements.  For small
displacements, the scattered intensity varies roughly as $Q^2$.  The reciprocal
lattice point $(\frac23,0,1)$ is at sufficiently small $Q$ that the
charge-order contribution is negligible.
\end{thebibliography}

\begin{figure}
\centerline{\psfig{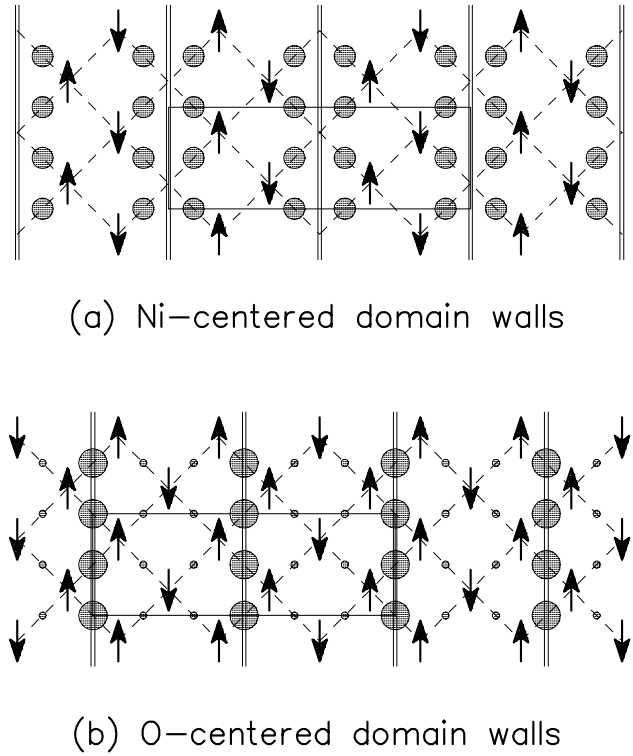}}
\bigskip
\caption{Stripe models for $\epsilon=\frac13$.  Arrows indicate correlated Ni
magnetic moments; shaded circles indicate locations of holes (on oxygen). 
Dashed lines trace the bonding paths of the square lattice, while solid lines
outline a unit cell. Double lines indicate positions of domain walls. (a)
Ni-centered (i.e., site-centered) domain walls.  All Ni moments (between domain
walls) are equivalent.  (b) O-centered (i.e., bond-centered) domain walls. 
Moments near domain walls and in the center of domains, respectively, are not
equivalent.
\label{fg:1}}
\end{figure}

\begin{figure}
\centerline{\psfig{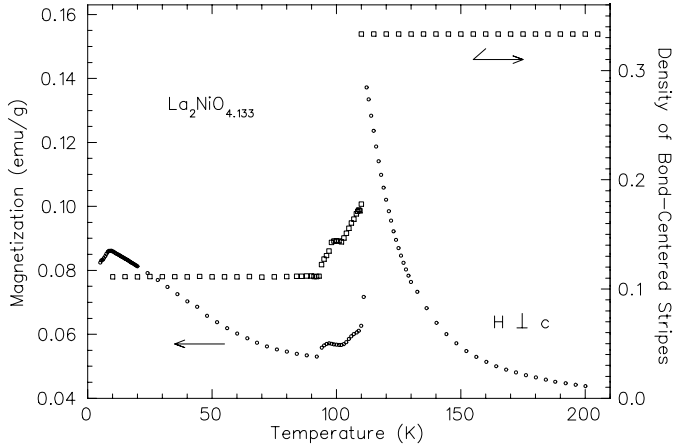}}
\bigskip
\caption{Bulk magnetization measured (on warming, after zero-field cooling) with
an applied field of 1~T aligned parallel to the NiO$_2$ planes (circles), and
density of bond-centered stripes (squares), which is equal to $4\epsilon-1$,
where values of $\epsilon$ are taken from Ref.~\protect\cite{woch96}.
\label{fg:2}}
\end{figure}

\begin{figure}
\centerline{\psfig{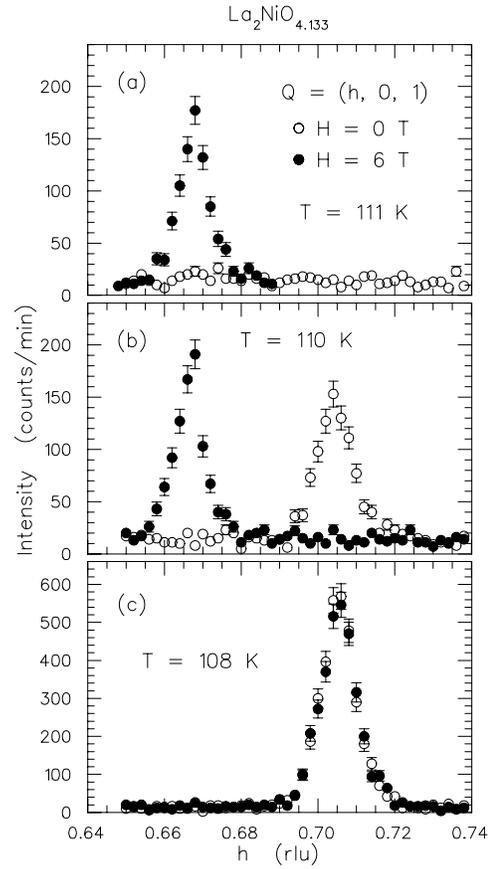}}
\bigskip
\caption{Elastic scans along ${\bf Q}=(h,0,1)$ in zero field (open circles) and
in a magnetic field of 6~T (filled circles).  Scans were measured at
temperatures of 111~K (a), 110~K (b), and 108~K (c).
\label{fg:3}}
\end{figure}

\begin{figure}
\centerline{\psfig{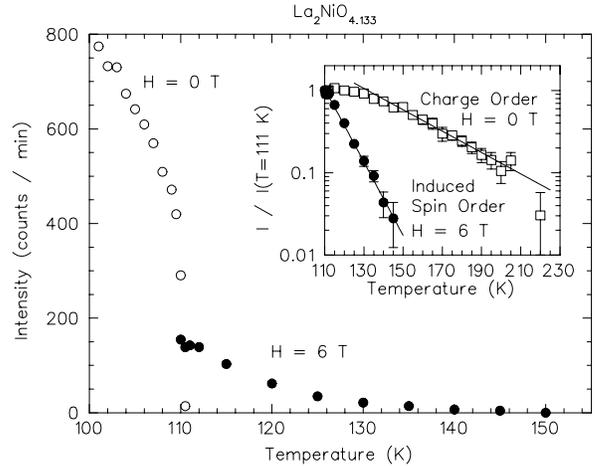}}
\bigskip
\caption{Temperature dependence of the intensity of the magnetic peak at ${\bf
Q}=(1-\epsilon,0,1)$ measured with $H=0$ (open circles) and $H=6$~T (filled
circles).  Below $\sim109$~K the magnetic peaks become independent of field. 
Inset: logarithm of normalized intensity vs.\ temperature for the magnetic peak
at $H=6T$ (filled circles), and for the charge order peak at
$(4-2\epsilon,0,1)$ (open squares, from Ref.~\protect\cite{woch96}). Lines
through points are linear fits, as discussed in the text.
\label{fg:4}}
\end{figure}

\end{document}